\documentclass[12pt]{iopart}
\usepackage{hyperref}
\usepackage{amsfonts}

\begin{document}

\title{Kolmogorov-Sinai entropy and black holes}\footnote{This is an extended version of my paper
arXiv:0711.3131.}
\author{Kostyantyn Ropotenko}
\address{State Department of communications, Ministry of transport and communications of
Ukraine, 22, Khreschatyk, 01001, Kyiv, Ukraine}

\begin{abstract}
It is shown that stringy matter near the event horizon of a
Schwarzschild black hole exhibits chaotic behavior (the spreading
effect) which can be characterized by the Kolmogorov-Sinai entropy.
It is found that the Kolmogorov-Sinai entropy of a spreading string
equals to the half of the inverse gravitational radius of the black
hole. But the KS entropy is the same for all objects collapsing into
the black hole. The nature of this universality is that the KS
entropy possesses the main property of temperature: it is the same
for all bodies in thermal equilibrium with the black hole. The
Kolmogorov-Sinai entropy measures the rate at which information
about the string is lost as it spreads over the horizon. It is
argued that it is the maximum rate allowed by quantum theory. A
possible relation between the Kolmogorov-Sinai and
Bekenstein-Hawking entropies is discussed.
\end{abstract}

\pacs{04.70.Dy} 

\bigskip\bigskip
\maketitle

\section{Introduction}

It is well known that some general relativistic systems described by
the Einstein equations can exhibit chaotic behavior \cite{bar,hob}.
But it is still not clear whether such general relativistic systems
as black holes can do it. In series of insightful papers, Susskind
\cite{s1,s2,s3} and Mezhlumian, Peet and Thorlacius \cite{mez}
showed that a string falling toward a black hole spreads over the
stretched horizon. In doing so, it exhibits chaotic behavior and the
rate of spreading is determined by the inverse gravitational radius
of the black hole.

In this paper I propose to use the conceptions of chaos theory to
describe the behavior of stringy matter near the event horizon of a
Schwarzschild black hole. In the following sections we will
introduce the main conceptions of chaotic dynamics, demonstrate
chaotic behavior of stringy matter near the event horizon of a black
hole, and find its Kolmogorov-Sinai entropy. In the end we discuss a
possible relation between the Kolmogorov-Sinai (KS) and
Bekenstein-Hawking (BH) entropies. This paper is an extended version
of \cite{ro}.

\section{The main conceptions of chaotic dynamics}

We begin with definitions. According to chaos theory \cite{zas,dor}
the chaotic behavior of dynamical systems that is, systems whose
state evolves with time, has its origin in the so-called local
instability, when a small change in initial conditions leads to an
exponential divergence of phase space trajectories. As a result of
this sensitivity to initial conditions, the behavior of dynamical
systems appears unpredictable. Suppose that two nearby trajectories
in the phase space of a system start off with a separation $d(0)$ at
time $t=0$. Then, if there exists a local instability, it follows
that there exists a direction in phase space along which the
trajectories diverge exponentially:
\begin{equation}
\label{tr} d(t)=d(0)\:e^{ht}.
\end{equation}
The parameter $h$ is called the Lyapunov exponent for the
trajectories. If $h$ is positive, then we say the behavior of the
system is chaotic. The sum of all the positive Lyapunov exponents is
called the KS entropy $h_{KS}$. It describes the rate of change of
information about the phase space trajectories as a system evolves.
Following Zaslavsky \cite{zas}, we can define the KS entropy in not
so formal way and connect it with thermodynamical entropy. Suppose
that the phase space of a dynamical system is finite (in at least
one phase space direction). Then, since there exists the local
instability (\ref{tr}), it follows that an initial volume in phase
space $\Delta{\Gamma(0)}$  becomes very complicated like a fractal.
Thus we should perform coarse graining, and if we coarse grain the
region, it will appear that it is growing
\begin{equation}
\label{ph} \Delta{\Gamma(t)}=\Delta{\Gamma(0)}\:e^{ht},
\end{equation}
where $h$ is the averaged over phase space Lyapunov exponent. As a
result, the coarse-grained Boltzmann entropy increases
\begin{equation}
\label{enB} S(t)=ht.
\end{equation}
The quantity $\Delta{\Gamma(0)}$ can be taken as simply equal to a
volume of the coarse-graining. Then the expression
\begin{equation}
\label{enkol}h_{KS} = \lim_{t\rightarrow
\infty}\lim_{\Delta{\Gamma(0)}\rightarrow 0}\frac{1}{t}\ln
\left(\frac{\Delta{\Gamma(t)}}{\Delta{\Gamma(0)}}\right)
\end{equation}
just defines the KS entropy. It is clear that the KS entropy
$h_{KS}$ is not really an entropy but an entropy per unit time, or
entropy rate, $dS/dt$. Note also that by order of magnitude
$h_{KS}\sim h$. As is easily seen, if $\Delta{\Gamma(0)}\rightarrow
0$ that is, if $S(t)=ht$ at $t\rightarrow\infty$, the Boltzmann
entropy doesn't reach a maximum on  a compact phase space. But the
situation changes if we fix a finite volume of coarse-graining. In
this case the exponential separation of trajectories (\ref{tr}) and
the consequent increase in volume (\ref{ph}) can last only on finite
time-scale. Just that very case will be discussed below. In general,
the calculation of the KS entropy offers great mathematical
difficulties. Our task is facilitated by the fact that all our
relevant quantities have exponential dependence on time.

The BH entropy of a black hole, on the other hand,
\begin{equation}
\label{BH}S_{BH}=\frac{A}{4G}=\frac{\pi R_{g}^{2}}{l_{P}^{2}}= 4\pi
GM^{2},
\end{equation}
is obtained from the thermodynamical relation $dE=TdS$, where the
energy of the black hole is its mass $M$, the temperature is given
by $T_H=1/8\pi GM$ and the area of the event horizon $A$ is related
to the gravitational radius $R_{g}$, $R_{g}=2GM$, in the usual way
$A=4\pi R_{g}^{2}$. The BH entropy is defined in the reference frame
of an external distant observer at fixed static position above the
horizon (an external observer). It is well established that, from
her/his point of view, the classical physics of a quasistationary
black hole can be described in terms of a 'stretched horizon' which
is a membrane placed near the event horizon and endowed with certain
mechanical, electrical and thermal properties \cite{tor}. The exact
distance of this membrane above the event horizon is somewhat
arbitrary. In the context of string theory - the subject of our
research - the stretched horizon is most naturally thought of as
lying at the string scale above the event horizon. In what follows
we will deal with this conception (sometimes referred to simply as
horizon).

It may seem that the event horizon itself exhibits the desired
instability: any light ray which is just not on the horizon
separates from it exponentially. This is very clear in
Eddington-Finkelstein coordinates. But it is not the local
instability in the phase space (\ref{tr}). Our purpose is to find a
kinematic effect caused by the black hole geometry with respect to
which a system evolves like (\ref{tr}), (\ref{ph}). For this purpose
we repeat, for completeness, some well-known facts from \cite{s4}
concerning the behavior of matter near the horizon without proofs,
thus making our exposition self-contained.

\section{Chaotic behavior of a relativistic string collapsing into a black hole}

My proposal rests on stringy matter having unusual kinematic
properties near the event horizon of a black hole. According to
string theory \cite{zwie}, the most promising candidate for a
fundamental theory of matter, all particles are excitations of a
one-dimensional object - a string. String theory is characterized by
two fundamental parameters: the string scale $l_{s}$, and the string
coupling constant $g$; if $l_{P}$ is the Planck length then
$l_{P}=g\: l_{s}$. An important fact is that behavior of a string in
weakly coupled string theory is very precisely described in terms of
the random-walk model \cite{zwie}-\cite{dam}. We can imagine a
string as simply built by joining together bits of string, each of
which is of length $l_s$. Suppose that the total length of the
string is $L$ and each string bit can point in any of $n$ possible
directions. Then the number of bits is $L/l_s$ and the number of
states of the string is
\begin{equation}
\label{str1} N_s \sim n^{L/l_s}\sim \exp {(L\ln n/l_s}),
\end{equation}
For notational simplicity the factor $\ln n $ will be omitted
henceforward. There is no loss of generality in doing so because we
can always redefine $g$, $l_S$ and $L$ . We can also define the mean
squared radius of the string $\langle R_s^{2}\rangle = Ll_s$. Then
\begin{equation}
\label{str2} N_s  \sim \exp {(\langle R_s^{2}\rangle/l_s^{2})},
\end{equation}
and the entropy of the string is
\begin{equation}
\label{str3} S_s \sim \frac{L}{l_s } \sim \frac{\langle
R_s^{2}\rangle}{l_s^{2}}.
\end{equation}
The mass of the string is given by $M_s\sim TL$, where $T\sim
1/l_s^{2}$ is the string tension. So in terms of the mass
\begin{equation}
\label{str4} S_s \sim M_s l_s.
\end{equation}
Note that in fundamental contrast to the black hole entropy
(\ref{BH}), the entropy of a string goes like the mass.

Another important fact is that strings behave very differently from
the ordinary particles. The crucial difference is that the size and
shape of a string are sensitive to the time resolution. It is a
smearing time over which the internal motions of the string are
averaged.  Susskind showed \cite{s1} that zero-point fluctuations of
a string make the size of the string depend on a time resolution;
the shorter the time over which the oscillations of a string are
averaged the larger is its spatial extent. In low energy physics,
resolutions times are always large and this phenomenon is not
important. Let, for example, an observation of a string lasts a time
$\tau _{res}$. Then Susskind argued that the contributions of modes
with frequency $\gg 1/\tau _{res}$ in the normal mode expansion for
the size of the string should be averaged out. As a result of such a
coarse-graining or time-smoothing, Susskind found that the mean
squared radii of the string in the transverse and longitudinal
directions in Planck units are $\langle R_{\perp}^{2}\rangle
=l_s^{2}\ln (1/ \tau _{res})$ and $\langle R_{\parallel}^{2}\rangle
=l_s^{2}/\tau _{res}$, respectively. Similar calculations can be
performed for the total length of the string, and Susskind found
that $L=l_s /\tau _{res}$. Now consider a string falling toward a
black hole. As is well known \cite{s4}, the proper time in the frame
of the string $\tau$ and the Schwarzschild time of an external
observer $t$ are related through $\tau \sim \exp({-t/2R_{g}})$  due
to the redshift factor. This means that the transverse size of the
string will increase linearly:
\begin{equation}
\label{str5}\langle R_{\perp}^{2}\rangle =l_s^{2}\frac{t}{2R_{g}},
\end{equation}
while its longitudinal size and total length - exponentially:
$\langle R_{\parallel}^{2}\rangle =l_s^{2}\exp({t/2R_{g}})$, $L
=l_s\exp({t/2R_{g}})$. But the longitudinal growth is rapidly
canceled by the Lorentz longitudinal contraction. Thus the string
approaching the horizon spreads only in the transverse directions
(in this connection the subscript '$\perp$' at the the mean squared
radius of the string will be replaced by the 's' henceforward).

In (\ref{str5}) we can immediately recognize the linear dependence
of the squared displacement of a Brownian particle from the origin
on time. The theory of Brownian motion is closely related to that of
random walks. One normally associates diffusion with the Brownian
motion of a particle. The Schr\"{o}dinger equation describing the
string has the same mathematical structure as the diffusion equation
of a Brownian particle. In fact the spreading appears to behave as
if the string was diffusing away from its original transverse
location. As is well known, the Brownian motion is a chaotic process
(moreover, it appears that one can even infers a positive KS entropy
from the Brownian motion (see \S 19.9 in Dorfman \cite{dor} and
references therein)). Thus a spreading string exhibits chaotic
behavior. Indeed, Susskind \cite{s1,s2,s3} and Mezhlumian, Peet and
Thorlacius \cite{mez} showed that the string configuration becomes
chaotic and very complicated like a fractal during the spreading
process (especially it is clear with the help of the Monte Carlo
simulation of the probability functional of a string \cite{s5}). As
mentioned above, during the spreading the total length of a string
increases exponentially that is, the number of bits increases
exponentially too:
\begin{equation}
\label{bit}N_{bit}=N_{bit}(0)\exp (t/2R_{g}).
\end{equation}
The authors interpreted this as a branching diffusion process, where
every string bit diffuses independently of others over the whole
horizon and bifurcates into two bits and so on. In this connection
it is relevant to remark the following. First the diffusion is a
distinctive random process. But in our case there are no real random
forces. The behavior of a string near the horizon is very well
described by the Hamilton dynamics. If there are exact equations of
motion no true randomness is possible. Second the string is a
fundamental object. It is not a dissipative system. In the spreading
process no points of a string should be lost or gained. The
irreversible character of the spreading effect arises, as has been
shown above, exceptionally from the coarse-grained or time-smoothed
procedure, in which the fine details of the string motion (over the
Planck time scale) have been wiped out. As an ordinary classical
body, a string undergos Lorentz transformations. As is well known
\cite{paul}, under the Lorentz transformations the phase space
volume of a classical body and, consequently, its entropy does not
change. But due to the coarse-grained procedure the phase space
volume of a string can increase. We can give the following
interpretation of (\ref{bit}). Initially phase points of a string
occupy one cell in the phase space of the string. In the course of
time, the number of bits increases. This means an increase in the
number of phase space dimensions. So the phase space volume
increases and we can interpret (\ref{bit}) as an increase in the
number of occupied cells:
\begin{equation}
\label{cel}N_{cell}(t) = N_{cell}(0)\exp (t/2R_{g}).
\end{equation}
In turn this number is  proportional to the distance between the
trajectories of phase points that all initially occupy one cell
(\ref{tr}), as required.

As we have seen, the growth of string (\ref{str5}) is linear. But as
noted by Susskind himself \cite{s3,s4}, this result was obtained in
the framework of free string theory. It doesn't take into account
such a nonperturbative phenomenon as string interactions; there are
indications \cite{s3,s4} that a true growth must be exponential:
\begin{equation}
\label{exp}\langle R_s ^{2}\rangle =l_s^{2}\exp({t/2R_{g}}).
\end{equation}
How does the phase space of string evolve during the spreading
process? The spreading process begins to occur when the string
reaches the horizon at distance of order of the string scale $l_{s}$
from the horizon in a thin layer $\sim l_{s}$. But in string theory
at such scales the mirror symmetry should takes place \cite{pol,
mar}. In general it relates the complex and K\"{a}hler structures of
some manifolds. In the simplest case for closed strings it exchanges
the winding number around some circle with the corresponding
momentum quantum number (T-duality) or, roughly speaking,
coordinates with momenta. At the scales $\gg l_{s}$ we can always
single out the configuration space and the phase space is its
cotangent bundle. At the scales $\sim l_{s}$ this is not the case:
at such scales there is a replacement of the configuration space of
a string by its phase space \cite{mar}. A similar phenomenon in
quantum mechanics - a particle in magnetic field \cite{lan}: on the
distances of order of the magnetic length $l_{mag} \sim \sqrt{\hbar
c/eH}$ a replacement of the configuration plane transversal to the
direction of the magnetic field by the phase plane takes place so
that the number of states is $A/l^{2}_{mag}$, where $A$ is the area
of the transversal plane. A string is not a point-like particle and
has its own phase space which can be very complex. But in any case
during the spreading process the configuration space of string
increases exponentially (in transverse directions). Taking into
account the replacement of the configuration space of a string by
its phase space we can conclude that the phase space of string will
increase exponentially in the corresponding directions:
\begin{equation}
\label{phstr} \Delta{\Gamma_s(t)}=\Delta{\Gamma_s(0)}\exp(t/2R_{g}).
\end{equation}
It is also true in the framework of the random-walk model. Since the
size of string increases exponentially, the relative velocities of
string bits should obey the Hubble law. It is obvious that in this
case the velocity subspace does not decrease and as a result the
whole phase space increases exponentially. We can also estimate the
number of states of a spreading string with the help of the
random-walk model. For example, in the case of the linear growth,
substituting (\ref{str5}) in (\ref{str2}), we obtain
\begin{equation}
\label{str6} N_s  \sim \exp {(\langle R_s^{2}\rangle/l_s^{2})}\sim
\exp(t/2R_{g}).
\end{equation}
So the entropy is
\begin{equation}
\label{enstr} S_s \sim \frac{t}{2R_{g}},
\end{equation}
and the entropy rate
\begin{equation}
\label{rate1} \frac{dS_s}{dt}=\frac{1}{2R_{g}}.
\end{equation}
But for the exponential growth of the size the same formulas give
$N_s \sim \exp(\exp (t/2R_{g}))$ (the same result is obtained and
for the total length). In this case the entropy $S_s \sim \exp
({t/2R_{g}})$ and the entropy rate
\begin{equation}
\label{rate2} \frac{dS_s}{dt}= \frac{S_s}{2R_{g}}.
\end{equation}
This is a comparatively large rate. The point is that we have used
the simple random-walk model which is valid only in the framework of
the free string theory. Apparently the interactions will impose
constraints on the total number of string states and the huge
entropy rate (\ref{rate2}) should reduce. But the precise
calculation of the entropy rate in the nonperturbative regime is
beyond the current technology of string theory. Despite this
difficulty we can estimate an upper limit on the entropy rate. As is
well known, the maximum rate at which information $I$ may be
transmitted at a temperature $T$ in quantum theory is given by
Pendry's formula \cite{pen,bek}:
\begin{equation}
\label{pen1} \frac{dI}{dt}\sim T.
\end{equation}
So the maximum rate is determined by temperature. Taking into
account the fundamental equivalence relation between information and
entropy we can rewrite (\ref{pen1}) in terms of entropy
\cite{pen,bek}:
\begin{equation}
\label{pen2} \frac{dS}{dt}\sim T.
\end{equation}
This expression can be immediately applied to the spreading process
near the horizon of a black hole with the Hawking temperature
$T_H=1/4\pi R_g$. Thus we obtain
\begin{equation}
\label{rate3} \frac{dS_s}{dt}\leq \frac{1}{4\pi R_g}.
\end{equation}
This is the maximum rate at which entropy of string can increase
during the spreading process. Up to a factor $1/2\pi$ it coincides
with the entropy rate for the linear growth (\ref{rate1}). It is
obvious that the entropy rate for the exponential spreading is
greater or equal to (\ref{rate1}) but less or equal to
(\ref{rate3}). So for the exponential spreading we obtain
\begin{equation}
\label{rate4} \frac{dS_s}{dt}=\frac{1}{2R_{g}}.
\end{equation}
Strictly speaking, the rate at which entropy of string (or any body
made of strings) increases during the spreading process saturates
the bound (\ref{rate3}).

Now, summarazing our arguments and comparing (\ref{phstr}),
(\ref{rate1}) with (\ref{ph}), (\ref{enB}) we can conclude that
stringy matter near the event horizon exhibits instability chaotic
behavior which can be characterized by the Kolmogorov-Sinai entropy
\begin{equation}
\label{KS1} h_{KS}=\frac{1}{2R_{g}}.
\end{equation}
Obviously the KS entropy is the same for all objects collapsing into
a black hole. So, it is universal. The nature of this universality
is that the KS entropy possesses the main property of temperature:
it is the same for all bodies in thermal equilibrium with a black
hole at the Hawking temperature $T_H=1/4 \pi R_g$.

It is relevant to remark that the behavior of string mentioned above
just corresponds to the thermal properties of a black hole and the
second law of thermodynamics. Since the temperature of the black
hole radiation depends on the radial position, $T(r)=T_H/\chi$,
where $\chi$ is the the redshift factor, $\chi=(1-R_g/r)^{1/2}$, it
follows that from the viewpoint of the external observer the string
falls into an increasingly hot region. A thermal interchange will
take place. So the string should 'melt' and spread throughout the
horizon. Obviously during this process the phase volume and entropy
of the body should increase. Thus (\ref{str6}), (\ref{enstr}) is a
natural response to the hot horizon.

How does the KS entropy reach its maximum? As mentioned above, there
is no point to define the configuration space of string with
accuracy better than $l_{s}$. The surface of the horizon is a
compact manifold. Since there exists the finite size of
coarse-graining $l_s$ (note that in strong coupling regime $l_s\sim
l_P$), a string covers the horizon in a finite time \cite{s3}
\begin{equation}
\label{time}t_{spread}\sim R_{g}\ln\frac{R_{g}^{2}}{l_P^{2}}.
\end{equation}
At this time a spreading string completely covers the entire horizon
of a black hole. The number of states of the string becomes $N\sim
\exp(R_g^{2}/l_P^{2})$ and the entropy of the string reaches that of
the black hole, $S\sim R_g^{2}/l_P^{2}$. Note that in doing so the
total length of the string becomes $L\sim R_g^{2}/l_P$ and the
corresponding mass $M_s\sim R_g^{2}/l_P^{3}$. It is a huge mass. But
the redshift factor reduces it to the black hole mass. Since the
spreading process begins to occur when the string reaches the
stretched horizon at the proper distance $l_P$ from the event
horizon, the redshift factor is $\chi\approx l_P/2R_g$, and we
obtain for the string mass $M_s\sim \chi (R_g^{2}/l_P^{3})\sim
R_g/(2l_P^{2})\sim R_g/2G$. Thus, from the string theory point of
view \cite{s4}, a black hole is nothing but a single string.
According to Susskind \cite{s1,s2,s3} and Mezhlumian, Peet and
Thorlacius \cite{mez}, at the time $t_{spread}$ the string spreads
over the entire horizon and can no longer expand due to the
nonperturbative effects. The result is crucial for the relaxation of
the string to statistical equilibrium: to reach a statistical
equilibrium in a finite time we should have the finite time of
spreading. According to chaos theory, this is an average time over
which the state of a string can be predicted; after the time
$t_{spread}$ all information contained in the string will be lost
and we will able only to make statistical predictions. This time is
comparable to the characteristic time of a black hole $R_{g}$ but is
smaller than the black hole lifetime $\sim R_{g}^{3}$. Hence the KS
entropy of a spreading string measures the rate at which information
about the state of the string (or any body made of strings)
collapsing into the black hole is lost with time as it spreads over
the horizon.

We have found the KS entropy for a fundamental string spreading over
the event horizon of a black hole. It is widely believed, however,
that the spreading effect is not a peculiar feature of a special
(still hypothetical) kind of matter. In the framework of the
so-called infrared/ultraviolet connection \cite{s4} it is a general
property of all matter at energies above the Planck scale. If we
want to study progressively smaller and smaller objects we must,
according to conventional physics, use higher and higher energy
probes. But once gravity is involved that rule is changed radically.
Since at energies above the Planck scale black holes are created, it
follows \cite{s4} that as we raise the energy we probe larger and
larger distances. In other words a very high frequency is related to
large size scale, $\Delta x \Delta \tau \sim l_{P}^{2}$. Then,
taking into account the redshift factor, we can obtain the
exponential growth of the transverse size of matter similar to
(\ref{str5}), as required.

\section{Is there a relation between the KS and BH entropies?}

As mentioned above, a falling string spreads over the stretched
horizon until its entropy becomes equal to the black hole entropy
\begin{equation}
\label{KS3} S_s = S_{BH}=\frac{\pi R_{g}^{2}}{l_{P}^{2}}.
\end{equation}
The spreading ends and only a new falling string or any other
perturbation can start it again. The next string falling toward the
horizon interacts with a previous one lying on the horizon in such a
way that the formation of a single (new) string is thermodynamically
favored, etc. So the stretched horizon is a single string made out
of all strings whenever fallen into it. From the string theory point
of view \cite{s4}, a black hole is nothing but a single string lying
on the sphere of the radius $R_g$. But since a black hole absorbs a
string, its gravitational radius must increase. According to the
teleological nature of the event horizon (see Chapter VI in Thorne
\emph{et al} \cite{tor}), before a fall of the next string, the
gravitational radius and the horizon area increase like $\exp
{(t/2R_{g})}$. This means that the spreading takes place. In this
case the entropy rate is
\begin{equation}
\label{k3} \frac{d \ln S_{BH}}{dt} =\frac{1}{2R_{g}}.
\end{equation}
It coincides with (\ref{rate2}) and does not obey the bound
(\ref{rate3}). To avoid this, we could assume that a true rate is
given by (\ref{rate4}). It seems, however, that it can still satisfy
Lloyd's limit. According to Lloyd \cite{llo}, the maximum rate at
which information can be moved in and out of a system with size $R$
and entropy $S$ is $dI/dt \sim S/R$ (attained by taking all the
information $S \ln 2$ in the system and moving it outward at the
speed of light). In my opinion, this aspect deserves further
investigation. In any case, the KS entropy of a black hole measures
the rate at which information about the black hole (or a string
forming the stretched horizon) is lost during a perturbation.

\section{Conclusions}

In this paper we have shown that stringy matter near the event
horizon of a black hole with the gravitational radius $R_{g}$
exhibits instability (the spreading effect) and chaotic behavior
which can be characterized by the Kolmogorov-Sinai entropy $h_{KS}$.
We have found that for a spreading string $h_{KS} = 1/2R_{g}$. But
it is the same for all objects collapsing into the black hole. It is
universal. The nature of this universality is that the KS entropy
possesses the main property of temperature: it is the same for all
bodies in thermal equilibrium with the black hole at the Hawking
temperature $T_H=1/4 \pi R_g$. The KS entropy of a spreading string
measures the rate at which information about the string (or any body
made of strings) collapsing into a black hole is lost as the string
(the body) spreads over the horizon. It is argued that it is the
maximum rate allowed by quantum theory. We have also discussed a
possible relation between the Kolmogorov-Sinai and
Bekenstein-Hawking entropies and suggested that $h_{KS}=d \ln
S_{BH}/dt$.

\section*{References}
\bibliographystyle{iopart-num}

\end{document}